\documentclass[aps,superscriptaddress,twoside,twocolumn,floatfix,a4paper,longbibliography,prb]{revtex4-1}

\usepackage{graphicx,epsfig}
\usepackage{color}
\usepackage[usenames,dvipsnames]{xcolor}
\usepackage[ocgcolorlinks,colorlinks=true,linkcolor=blue,citecolor=red]{hyperref}
\usepackage{amsmath,amssymb, amsthm, amsfonts}
\usepackage{dsfont} 
\usepackage{xspace}
\usepackage{xcolor}
\usepackage{changes}
\usepackage{verbatim}
\usepackage{mathtools}

\DeclareMathAlphabet{\mathbbold}{U}{bbold}{m}{n}

\newcommand{\bb}[0]{\begin{eqnarray}}
\newcommand{\ee}[0]{\end{eqnarray}}

\begin{document}

%\title{Phase-coherent heat engine based on a mesoscopic Aharonov-Bohm interferometer}
\title{Efficient and tunable Aharonov-Bohm quantum heat engine}

\author{G\'eraldine Haack}
\affiliation{Department of Applied Physics, Universit\'e de Gen\`eve, 1211 Gen\`eve, Switzerland}
\author{Francesco Giazotto}
\affiliation{NEST, Istituto Nanoscienze-CNR and Scuola Normale Superiore, Piazza San Silvestro 12, 56127 Pisa, Italy}

%\date{\today}  

\begin{abstract}
We propose a quantum heat engine based on an Aharonov-Bohm interferometer in a two-terminal geometry, 
and investigate its thermoelectric performances in the linear response regime. Sizeable thermopower (up to $\sim 0.3\,\text{mV}$/K) as well as $ZT$ values largely exceeding unity can be achieved by simply adjusting parameters of the setup and temperature bias across the interferometer leading to thermal efficiency at maximum power approaching $30\%$ of the Carnot limit. This is close to the optimal efficiency at maximum power achievable for a two-terminal heat engine. Changing the magnetic flux, the asymmetry of the structure, a side-gate bias voltage through a capacitively-coupled electrode and the transmission of the T-junctions connecting the AB ring to the contacts allows to finely tune the operation of the quantum heat engine. The exploration of the parameters' space demonstrates that the high performances of the Aharonov-Bohm two-terminal device as a quantum heat engine are stable over a wide range of temperatures and length imbalances, promising towards experimental realization. %Despite the simplicity of the setup, the high performances of the engine are stable over a wide range of temperatures and length imbalances, promising towards experimental realization.
\end{abstract}

\maketitle

\section{Introduction}
 
The investigation of thermal properties and heat transport at the nanoscale has garnered an impressive attention in the last few years \cite{Giazotto06, Binder, Fornieri17, Giazotto2012, Whitney18}. Strong advances achieved so far have proven a deeper understanding of the fundamental processes governing thermal transport and dynamics in solid-state nano-systems, both from the theoretical \cite{Giazotto02, Pekola07, Sothmann12, Jordan13, Ozaeta14, Giazotto14, Sanchez15, Hofer15, Giazotto15, Marchegiani16, De16, Samuelsson17, Hwang18, Kheradsoud19} and experimental side \cite{Savin04, Meschke06, Saira07, Giazotto12, Thierschmann15, Roche15, Josefsson18, Peterson18}. 
A very relevant question is related to establish if, and up to which extent, quantum effects do play a role in setting and controlling the performance of nano-sized heat engines, for instance, their conversion efficiency and  output power \cite{Houten92, Whitney, Benenti17}.

\begin{figure}[ht!]
%\begin{center}
\includegraphics[width=1.0 \columnwidth]{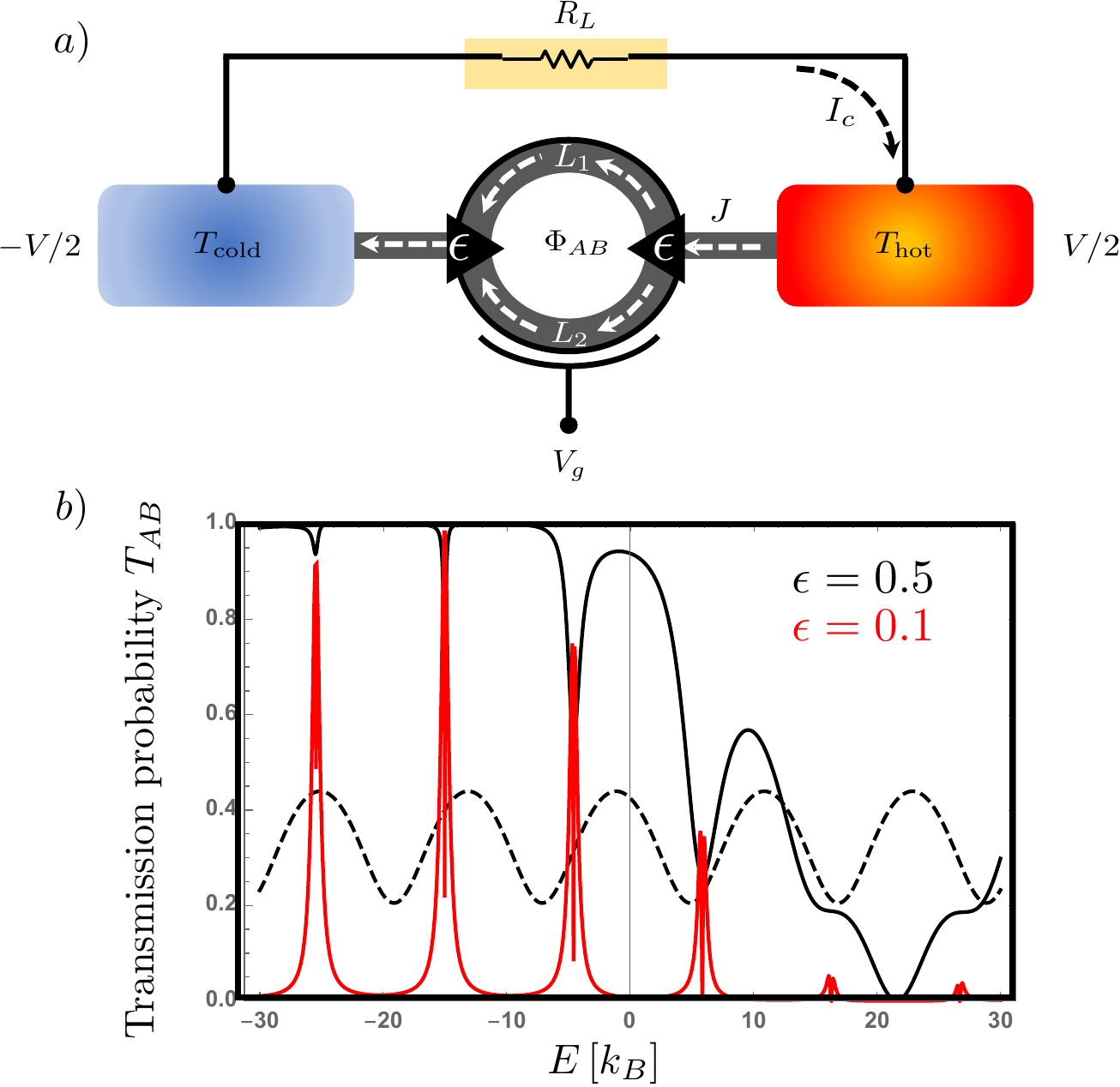}
\caption{a) Scheme of the two-terminal Aharonov-Bohm (AB) quantum heat engine. $L_1$ and $L_2$ denote the lengths of the arms of the interferometer whereas $T_{hot}$ and $T_{cold}$ are the temperatures of the hot and cold reservoirs. $\Phi_{AB}$ is the magnetic flux piercing the loop, $V_g$ is the gate voltage, and $J$ is the heat current flowing through the system. Transmission from the leads to the AB ring goes through T-junctions, parametrized by $\epsilon \in [0, 1/2 ]$.  The AB interferometer is connected to a generic load resistor $R_{L}$ through dissipationless (superconducting) wires. $V$ and $I_{c}$ represent the thermovoltage developed across the AB ring, and the thermocurrent circulating in the circuit, respectively. b) Tunability of the energy-dependent transmission probability $T_{AB}$ of the AB ring as a function of the asymmetry $\delta L = 0$ (dashed) and $\delta L/L = 0.3$ (solid) and as a function of the transmittivity of the T-junctions $\epsilon = 0.5$ (black) and $\epsilon = 0.1$ (red) for $e V_g = 2 \pi \mu/5 , 2 \pi \Phi_{AB}/\Phi_0 = \pi/7, L = 2 \cdot 10^{-6} \mu$m.}
\label{ring}
%\end{center}
\end{figure}

Albeit classical thermodynamics turns out to hold at the microscale, it is well established that quantum effects and \emph{phase coherence} may have a profound impact  on the overall behavior of nanoscopic heat engines. 
In the above context, the celebrated \emph{Aharonov-Bohm effect} \cite{Aharonov59}, i.e., the quantum mechanical property for which a charged particle can be influenced by either electric or magnetic potentials, may represent the prototypical building block for the realization of efficient quantum heat engines.

Here we envision and analyze a solid-state phase-coherent heat engine based on a mesoscopic Aharonov-Bohm (AB) interferometer, as shown in Fig. \ref{ring}.  Our proposal has a number of peculiar and attractive features: i) The working medium of the engine is a mesoscopic AB ring, that provides quantum phase-coherent control of particle and heat currents; ; ii) The system can operate as an ideal \emph{heat switch} by controlling very precisely the thermal conductance $\kappa_{th}$ from a thermally insulating to a conducting state. 
iii) The Seebeck coefficient $S$ (thermopower) can be large, i.e., of several hundreds of $\mu$V/K for suitable system parameters and temperature bias across the interferometer, and of both positive and negative sign whereas the dimensionless figure of merit $ZT$ obtains values largely exceeding unity; iv) The heat engine efficiency at maximum power approaches $30\%$ of the Carnot efficiency, which is close to the theoretical optimum \cite{Whitney}. 
The AB flux threading the sample plays the role of a finely tunable parameter, allowing for switching on and off the engine. Properties i)-iv) are optimized by controlling the asymmetry of the ring, a side gate voltage through a capacitively-coupled gate electrode and the quality of the contacts between leads and working medium. %As the engine working medium relies on single-particle interferences, a genuine quantum effect, the AB setup we propose  s is purely quantum, the AB setup is a clear example that quantum mechanics can be rather advantageous for the realization of efficient thermal machines. 
The AB interferometer represents the archetypal quantum platform in mesoscopic physics which has been largely investigated so far both from the theoretical \cite{Buttiker84, Yeyati95, Blanter97, Schoemerus07, Dolcini07} and experimental \cite{Webb85, Washburn86, Yacobi96, Yamamoto12} side since it can be easily realized with current state-of-the-art nano-fabrication techniques and available materials, based on metallic or 2DEG structures. Transport through this device takes place in 1D. In light of the above considerations, the AB engine has the potential to pave the avenue to the realization of a novel class of phase-tunable quantum machines ranging from thermal rectifiers \cite{Lopez13, Martinez13, Giazotto13, Martinez15, Sanchez17, Marcos18, Roche15} and autonomous heat engines \cite{Serra16, Roulet17, Tonekaboni18, Monsel18} to driven Floquet heat engines \cite{Arrachea07, Moskalets16, Dare16, Ludovico16a, Ludovico16b, Restrepo18}. \\

The paper is organized as follows. In Sec.~\ref{Sec:TAB}, we provide the expression of the transmission probability $T_{AB}$ of the AB ring using a scattering matrix approach, all details are given in App~\ref{App:A}. %Although the approach is not new and the derivation does not present any specific difficulty, the complete expression of $T_{AB}$ as a function of the length and asymmetry of the ring, in presence of a gate voltage and for finite transmitting T-junctions is not present in the literature. As we demonstrate with the AB ring can be operated as efficient thermoelectric device by tuning those parameters, we believe it is important for the community to provide the full expression of $T_{AB}$.
In Sec.~\ref{Sec:thermo}, we investigate the behaviour of the electrical and thermal conductances, as well as of the Seebeck coefficient, as a function of key experimental parameters, for instance, the average temperature, the gate voltage, the AB flux and the size and asymmetry of the ring. We show that the AB ring can be tuned to reach high thermopower, confirmed by values of the \emph{ZT} figure of merit largely exceeding 1. In Sec.~\ref{Sec:efficiency}, we exploit the good thermoelectric performances of the AB ring to propose a heat engine with an AB ring as working medium. By adjusting the value of the load resistance, we show that it achieves a close-to-optimal efficiency, $\eta \sim 30\%$. We conclude this work by investigating in Sec.~\ref{Sec:T_junction} the effect of non-ideal T-junctions, \textit{i.e.}, when incoming electrons from the contacts do have a finite probability to be reflected into the contacts. This leads to resonant tunneling, increasing even more the performance of the AB heat engine. % Although it does not affect neither the figure of merit nor the use of the AB ring as heat switch at low temperatures, it decreases the efficiency of the the heat engine.   
 
\section{Linear response regime \\
and transmission probability} 
\label{Sec:TAB}

In the linear response regime, which is valid when $T_\text{hot} - T_\text{cold} \equiv \Delta T \ll T \equiv (T_\text{hot} + T_\text{cold})/2$, and within a Landauer-B\"uttiker formalism, the charge and heat currents, respectively $I$ and $J$, are related to a voltage bias $\Delta V$ and temperature bias $\Delta T$ by the transport coefficients \cite{Houten92, Benenti17}:
\bb
\label{eq:Onsager}
\left(\begin{array}{c}
I \\
J \end{array} \right) = \left( \begin{array}{cc} 
G & L \\
M & K \end{array}\right) \left( \begin{array}{c}
\Delta V \\
\Delta T \end{array}\right)\,.
\ee
These coefficients take the explicit form $G = 2 e^2 I_0, L = 2 e I_1/T, K = 2 I_2/T$, and $M=2 e I_1 = L T$ with the integrals
\bb
\label{eq:I_n}
I_n &=& \frac{1}{h} \int_{-\infty}^{\infty} dE \, (E-\mu)^n\, T_{AB}(E) \left(- \frac{\partial f}{\partial E}\right) \,\, n=0,1,2\,. \nonumber \\
&& 
\ee
The factor 2 in the transport coefficients accounts for spin-degeneracy, and the Fermi distribution's derivative is $(-\partial f/\partial E) = \Big[4 k_B T \cosh^2[(E-\mu)/(2k_B T)]\Big]^{-1}$. The coefficients $G, L, K$ fully determine the thermal conductance $\kappa_{th}$ , as well as the thermo-electric response of the device, \textit{i.e.,} the Seebeck coefficient $S$
\bb
\kappa_{th} &=&  \left. \frac{J}{\Delta T}\right\vert_{I=0} = -(K + S^2 G T)  \label{eq:kappa}\\
S&=&  \left. \frac{\Delta V}{\Delta T}\right\vert_{I=0} = - \frac{L}{G}\,.  \label{eq:S}
\ee
The Seebeck coefficient depends on the ratio $I_1/I_0$ as defined in Eq. \eqref{eq:I_n}, which vanishes if the transmission probability $T(E)$ is symmetric with respect to energy or energy-independent. Below, we show that it is the combination of optimal values for the asymmetry of the ring and of the gate voltage essentially that leads to a strong energy-asymmetry in $T$, and that allows the AB ring to exhibit strong thermoelectric effects. To derive the transmission probability through the AB ring within a scattering-matrix approach, % Hence, it is behaviour of the transmission probability $T_{AB}(E)$ as a function of energy that will determine whether the AB ring can be exploited efficiently as thermoelectric device. 
we model our setup as a two-terminal geometry consisting of an AB ring connected to two metallic reservoirs through T-junctions. The interferometer is threaded by a magnetic field that leads to an enclosed magnetic flux $\Phi_{AB}$ and is gated via an external voltage $V_g$, see Fig. \ref{ring} a). Although the approach is not new, see in particular Refs.~\onlinecite{Buttiker84, Blanter97, Nazarov09}, and the derivation does not present any specific difficulty, the complete expression of $T_{AB}$ as a function of the length and asymmetry of the ring, in the presence of a gate voltage and for finite-transmissivity T-junctions is absent in the literature, to the best of our knowledge. 
Since we will demonstrate that the AB ring can be operated as an efficient thermoelectric device by tuning those parameters, we believe it is important  to provide the full expression of $T_{AB}$. The full derivation with all the details is given in App.~\ref{App:A}, the final expression being:

\begin{widetext}
 \bb
 \label{eq:TAB}
 T_{AB} &=& \frac{  1- \cos\chi \cos \delta \chi + \cos(2 \pi \Phi_{AB}/\Phi_0) \,(\cos \delta \chi  - \cos \chi)}{ \sin^2 \chi  + \left( \frac{2(1-\epsilon) \cos \chi  - \big(1-\epsilon - \sqrt{1-2 \epsilon}\big)\cos \delta \chi  -  \big(1-\epsilon + \sqrt{1-2 \epsilon}\big)\cos(2 \pi \Phi_{AB}/\Phi_0) }{2 \epsilon} \right) ^2 }. \nonumber \\
 &&
 \ee
 \end{widetext}
 
 \begin{figure*}[t!]
\label{fig:transport}
\begin{center}
\includegraphics[width=0.8 \paperwidth]{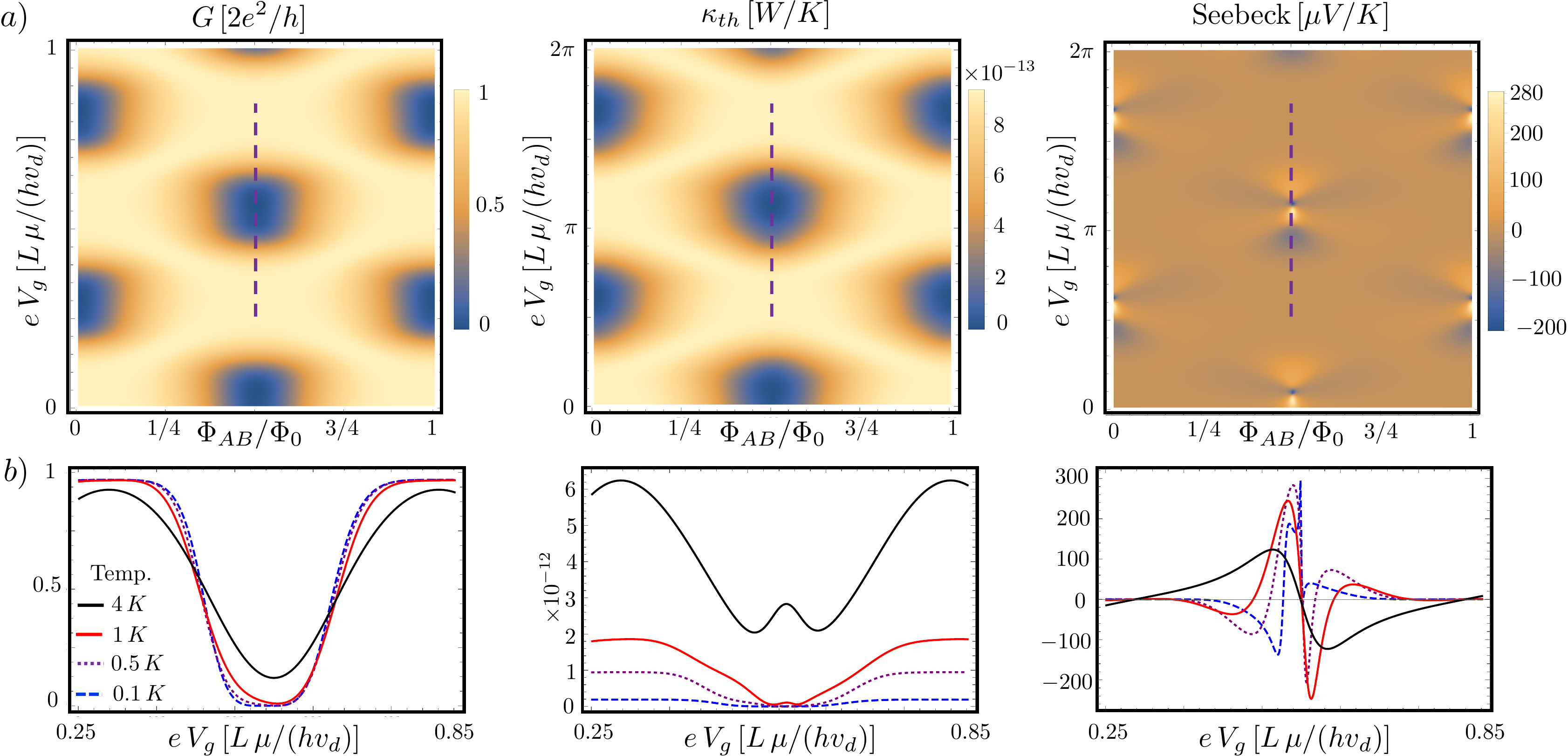}
\caption{a) Density plots of the electric ($G$) and thermal ($\kappa_{th}$) conductance, and Seebeck coefficient vs magnetic flux $\Phi_{AB}$ and gate voltage $V_g$. 
Length and imbalance of the ring are respectively fixed to $L= 2 \, \mu$m and $\delta L/L = 0.3$.
b) The same quantities as in  a) vs $V_g$ calculated along the dashed-line cut ($\Phi_{AB}/\Phi_0=1/2$) for a few selected temperatures. The electronic drift velocity is set to $10^{6} \text{m.s}^{-1}$, and $\tilde{k}_\mu = \sqrt{\pi} k_\mu$.}
\label{fig2}
\end{center}
\end{figure*}

%\begin{widetext}
%\bb
%\label{eq:TAB}
%T_{AB} =  \frac{1- \cos(\chi_1+\chi_2) \cos(\chi_1-\chi_2) + \cos(2 \pi \Phi_{AB}/\Phi_0) \Big( \cos(\chi_1-\chi_2) - \cos(\chi_1+\chi_2) \Big)}{\sin^2(\chi_1+\chi_2) + \left( \cos(\chi_1+\chi_2) - \frac{\cos(\chi_1-\chi_2) + \cos(2 \pi \Phi_{AB}/\Phi_0)}{2} \right)^2 }
%\ee
%\end{widetext}
Here, $\chi = \chi_1 + \chi_2$ and $\delta \chi = \chi_1 - \chi_2$, where $ \chi_i = k_i L_i$ are the dynamical phases that electrons acquire while traveling in each arm $i=1,2$, $k_i$ being the wave vector and $L_i$ the length of the corresponding arm. The AB flux is $2 \pi \Phi_{AB}/\Phi_0$ with the flux quantum $\Phi_0 = h/e$. 
Furthermore, non-ideal T-junctions, implying that  electrons arriving from source or drain contacts onto the beam splitter have a finite probability to be reflected back to the contacts, are characterized by a single parameter $\epsilon \in [0,1/2]$, see Ref. \onlinecite{Buttiker84}. 
This parameter allows to continuously span from a fully transmitting T-junction ($\epsilon = 1/2$) to a fully disconnected AB ring from the source and drain for $\epsilon = 0$. The energy-dependence of $T_{AB}$ is put in evidence by linearizing the spectrum around the Fermi energy $\mu$. Taking into account an additional voltage gate $V_g$ applied onto the lower arm and the possibility to tune the Fermi wave vector with some energy offsets applied onto each arm ($k_\mu \rightarrow \tilde{k}_\mu$), the wave vectors for electrons travelling through the upper and lower arms are respectively given by
\bb
k_1(E) &=& \tilde{k}_\mu + \frac{E-\mu}{\hbar v_d}\, , \,  k_2(E) =  \tilde{k}_\mu + \frac{E-(\mu+eV_g)}{\hbar v_d}\,.
\ee 
For an asymmetric AB ring, the difference and sum of the dynamical phases then read:
\bb
\label{eq:chi}
\!\!\!\!\chi_1 + \chi_2 &=& (2 L + \delta L) \left(\tilde{k}_\mu + \frac{E-\mu}{\hbar v_d} \right)  - \frac{e V_g L}{\hbar v_d} \label{eq:chis}\\
\!\!\!\!\chi_1 - \chi_2 &=& \delta L \left(\tilde{k}_\mu + \frac{E-\mu}{\hbar v_d} \right)  + \frac{e V_g L}{\hbar v_d}\,. \label{eq:chid}
\ee
Here we have defined $L_1 \equiv L + \delta L$, $L_2 \equiv L$, and $v_d$ is the electronic drift velocity. By inserting Eqs. \eqref{eq:chis} and \eqref{eq:chid} into Eq. \eqref{eq:TAB}, we obtain the energy-dependent transmission probability as a function of the arm's length $L$, the asymmetry of the ring $\delta L$, the transmission of the T-junction $\epsilon$, the AB flux $\Phi_{AB}$ and the gate voltage $V_g$. \\

From Eqs. \eqref{eq:TAB}, \eqref{eq:chis}, \eqref{eq:chid}, it becomes clear why the key parameters for enhancing the thermoelectric properties reduce to the length $L$, the asymmetry $\delta L$, the gate voltage $V_g$ and the transmission parameter of the T-junctions $\epsilon$. Indeed, as stated earlier, energy asymmetry is the key factor, and the energy dependence arises in the dynamical phases $\chi_1$ and $\chi_2$, which can be controlled by the afore-mentioned parameters. By contrast, the AB flux can be exploited to turn on and off the thermoelectric properties of the AB ring. Remarkably, when $\epsilon \ll 1$ corresponding to poorly transmitting T-junctions, the transmission probability $T_{AB} (E)$ exhibits resonance peaks as a function of energy. In other words, when the T-junctions are poorly transmitting, a \emph{resonant tunneling} effect through the AB ring is observed. Thanks to the complexity of the energy-dependence of $T_{AB}(E)$ as a function of the multiple parameters of the AB ring, we can exploit them to design a close-to-ideal transmission probability for maximising thermoelectric effects. This is achieved when the transmission probability is made of a series of delta-peaks (originating in the resonance peaks for $\epsilon \ll 1/2$), highly asymmetric with respect to the Fermi energy. As an illustration of those properties, Fig. \ref{ring} b) shows the strong energy asymmetry of the transmission probability in the presence of a finite asymmetry $\delta L$ and  external voltage $V_g$, for optimal ideal T-junctions ($\epsilon = 0.5$) and for poorly transmitting T-junctions ($\epsilon = 0.1$). %Without one of these specificities, the AB ring can not be considered as promising quantum heat engine. 

\section{Thermoelectric properties of the AB ring}
\label{Sec:thermo}

We first investigate the parameters' space of the AB ring to determine under which conditions the AB ring exhibits strong thermoelectric effect. To this end, we first assume that the T-junctions are fully transmitting [$\epsilon = 0.5$ in Eq. \eqref{eq:TAB}]. The effect of finite reflection at the T-junctions is investigated in Sec.~\ref{Sec:T_junction}. In Fig. \ref{fig2}, we show the transport and thermoelectric coefficients of the gated AB ring as a function of the AB flux and the gate voltage $V_g$ for a fixed length $L = 2 \, \mu \text{m}$ and length asymmetry $\delta L/L = 0.3$. The dependence on average temperature is shown in panel b), whereas the density plots in panel a) are valid for an average temperature $T= 500 \, \text{mK}$.

As expected, at low temperatures for which the Fermi distributions are almost step-like functions, the electrical conductance exhibits maximal amplitude $2 e^2/h$ since it is principally determined by the transmission probability which ranges from 0 to 1 over the parameter space. 
By contrast, as the temperature increases, we observe a decrease in its maximal amplitude.
 
The behaviour of the thermal conductance is of particular relevance for exploiting the AB ring to control heat currents. 
The density plot of $\kappa_{th}$ as a function of the gate voltage and AB flux shows that the latter can be used to fully control heat currents, i.e., the AB ring behaves as an ideal phase-tunable \emph{heat switch}. By adjusting the AB flux allows one to tune the thermal conductance from 0 to maximum. This type of device is highly desirable towards phase-coherent caloritronics  \cite{Fornieri17, Giazotto2012}.  
This operating mode has to be contrasted with existing proposals for quantum heat switches based on superconducting devices and Josephson junctions. 
There, the switch is only over the phase-dependent component (due to the Josephson coupling) of the thermal conductance, but there is always the phase-independent contribution in the background \cite{Giazotto2012}.  %the AB ring can be operated as an almost ideal phase-coherent heat switch for heat currents. 
As expected, and highlighted in panel b) for the thermal conductance, the AB-ring based quantum heat switch loses quality as temperature increases.

%In addition to the AB flux and the gate voltage $V_g$, we also investigate the transport and thermoelectric properties of the gated AB ring as a function of the average temperature $T$, and the length and imbalance of the ring, $L$ and $\delta L$.  Figure 2, panel a), shows the electrical and thermal conductances, as well as the Seebeck coefficient for a fixed length of $L = 2 \, \mu \text{m}$ and $\delta L/L = 0.3$ at temperature $T= 500 \, \text{mK}$. We note that both conductances show a similar behaviour as a function of the parameters. They reach maximal (minimal) values, indicating that the AB ring can also serve to fully control heat currents towards phase-coherent caloritronics (i.e., behaving as an ideal heat switch), using both $V_g$ and the AB flux \cite{Fornieri17}. 

\begin{figure}[t!]
\begin{center}
\includegraphics[width=0.8 \columnwidth]{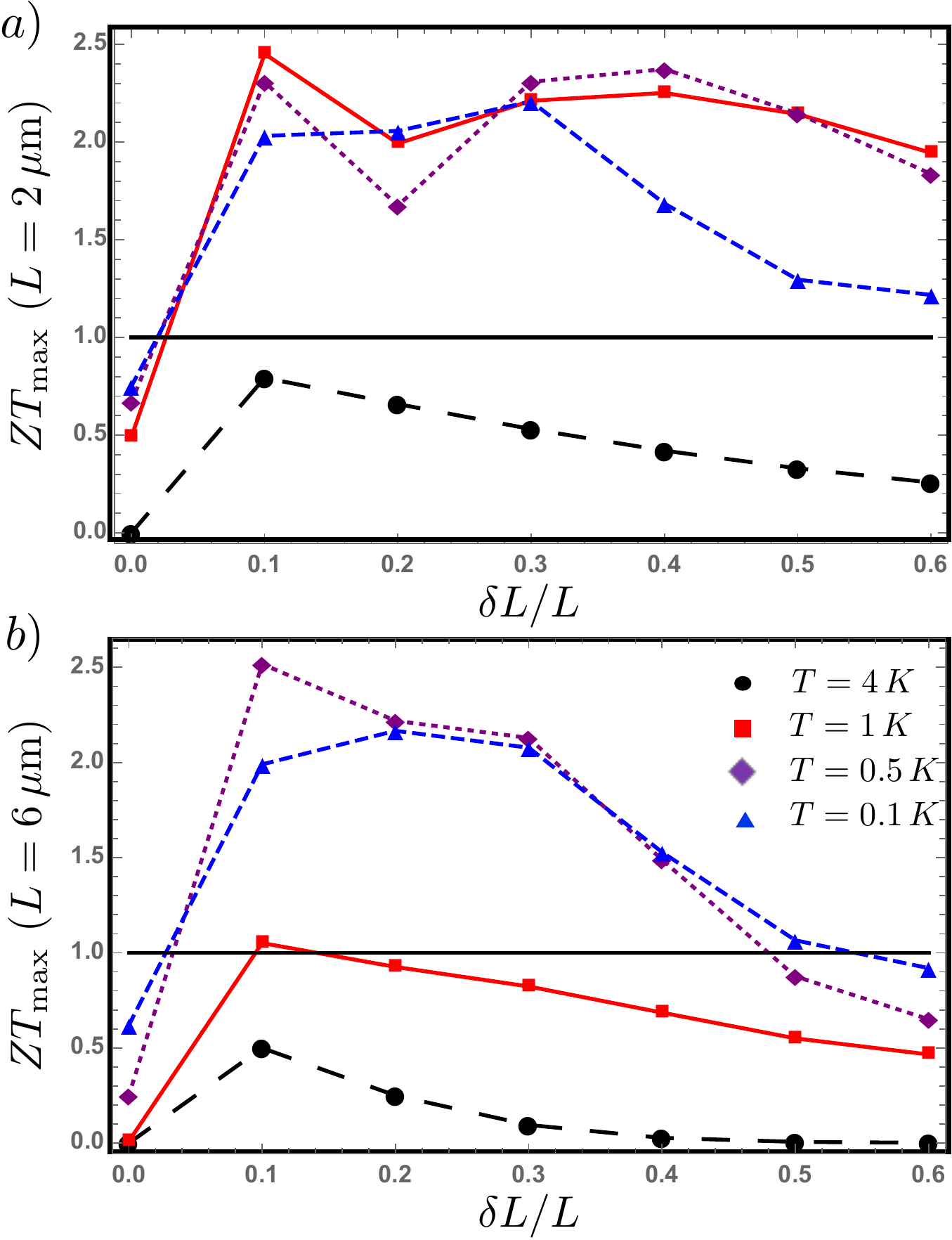}
\caption{Maximum $ZT$ coefficient for the AB heat engine ($ZT_{max}$) calculated as a function of ring asymmetry $\delta L$ at different temperatures $T= 0.1, 0.5, 1,4\,$K for a) $L=2\mu$m,  and b) $L=6\mu$m. For each asymmetry, the optimal $ZT$ is shown, reached for different values of the AB flux and gate voltage. 
$ZT_{max}$ largely exceeds 1, a critical value for an engine to be considered as a promising thermoelectric device. }
\label{fig:ZT}
\end{center}
\end{figure}

The thermoelectric response of the device is given by the Seebeck coefficient $S = V/\Delta T\vert_{I=0}$. 
It characterizes the amount of electrical voltage that is generated across the AB ring by a thermal bias, at zero electrical current. As a function of $V_g$, it exhibits a strong asymmetry, taking both positive and negative values. This reflects whether electron- or hole-like excitations contribute predominantly. Remarkably, it reaches values as high as 300 $\mu$V/K, twice the values obtained by operating a Mach-Zehnder interferometer as heat engine \cite{Hofer15}, and several times larger than earlier devices based on tunneling quantum dots \cite{Jordan13} and chaotic cavities \cite{Sothmann12}. As the temperature increases, the Seebeck coefficient also exhibits a smaller amplitude, and we therefore expect lower performance of the corresponding heat engine. 

In the linear response regime, a valid figure of merit for thermoelectric properties of a device is the \emph{ZT} coefficient, defined as $ZT = G S^2 T /\kappa_{th}$. It corresponds to the ratio of heat current originating from purely thermoelectric effects to the heat current in absence of a voltage bias \cite{Behnia}. Hence, desirable values of \emph{ZT} are above 1.
%Panel b) of Fig. 2 displays cuts of the same three quantities for a fixed AB flux, $\Phi_{AB}/\Phi_0 = 1/2$, as a function of $V_g$ for different temperatures. At high temperatures, $T=4$K, the electrical and thermal conductances do not exhibit maximal amplitude, preventing the  use of the AB ring to fully control heat current. The Seebeck coefficient also exhibits a smaller amplitude at $T=4$K, compared to lower temperatures, and we therefore expect lower performance of the corresponding heat engine. 
%This is confirmed by investigating the $ZT$ figure of merit, valid in the linear response regime. 
In Fig. \ref{fig:ZT}, we show the maximal $ZT$ values obtained over the full parameter space spanned by the AB flux and gate voltage as a function of the imbalance of the interferometer $\delta L/L$. All maximal values for the \emph{ZT} coefficient were obtained for AB flux equal to 0 or $1/2 \Phi_0$, only the gate voltage $V_g$ has to be adjusted when changing $L$ and $\delta L$. 
This specificity originates in the energy dependence of the transmission probability $T_{AB}$ that is controlled exclusively by $V_g, L $ and $\delta L$, as discussed in the previous section. This result is of particular interest to further exploit the AB ring as a cyclic engine, for instance, since it allows to turn on and off the thermoelectric properties of the AB ring with the magnetic  flux only.

%These maximal values were all obtained for fluxes equal to 0 or $1/2$, but for various values of the gate voltage. 
Both panels confirm that a finite imbalance is crucial to get $ZT$ values larger than one, and therefore to operate the gated AB ring as an efficient thermoelectric device. Since the dependence on the length $L$ of the transmission probability is highly non-linear, it is difficult to make general statements about \emph{ZT} values as a function of $L$ and $\delta L$. However, the good performances of this device are found for a wide range of length asymmetries, up to $\delta L / L =0.6$ for $L = 2 \mu$m, and up to 0.4 for for $L = 6 \mu$m.  It also appears that high \emph{ZT} values are found for average temperatures ranging from $0.5\,$K to $1\,$K, thereby demonstrating that good thermoelectric performances of the AB ring are not too sensitive to key experimental parameters. This is very promising towards a realistic implementation of the AB heat engine, and was not known until now. 
% are optimal to operate this mesoscopic interferometer. 
Figure \ref{fig:ZT} also indicates that the maximal $ZT$ values at $0.1\,$K decrease more rapidly when increasing asymmetry, and the maximal $ZT$ values at $4$K are all below the threshold value 1. 
Comparing now panels a) and b), it appears that the figure of merit of this engine decreases when increasing the length $L$ of the arms of the ring, for instance when going from $L=2\, \mu$m to $L= 6\,\mu$m. 
Let us now emphasize that the latest values of the electronic coherence length in AB rings are of the order of $l_\phi \simeq 6 \, \mu$m at $T=1\,$K, and $l_\phi \simeq12 \mu$m at $T=0.5\,$K \cite{Yamamoto12}. This experimental achievement, together with the relative ease of realization of this device with standard fabrication techniques in high-mobility GaAs-AlGaAs two-dimensional electron gas  heterostructures, makes the AB ring very promising for being a realistically efficient and highly-tunable quantum heat engine, whose functioning relies on single-particle quantum interferences.

\section{The AB ring as efficient quantum heat engine} 
\label{Sec:efficiency}

\begin{figure}[t!]
\begin{center}
\includegraphics[width=0.8 \columnwidth]{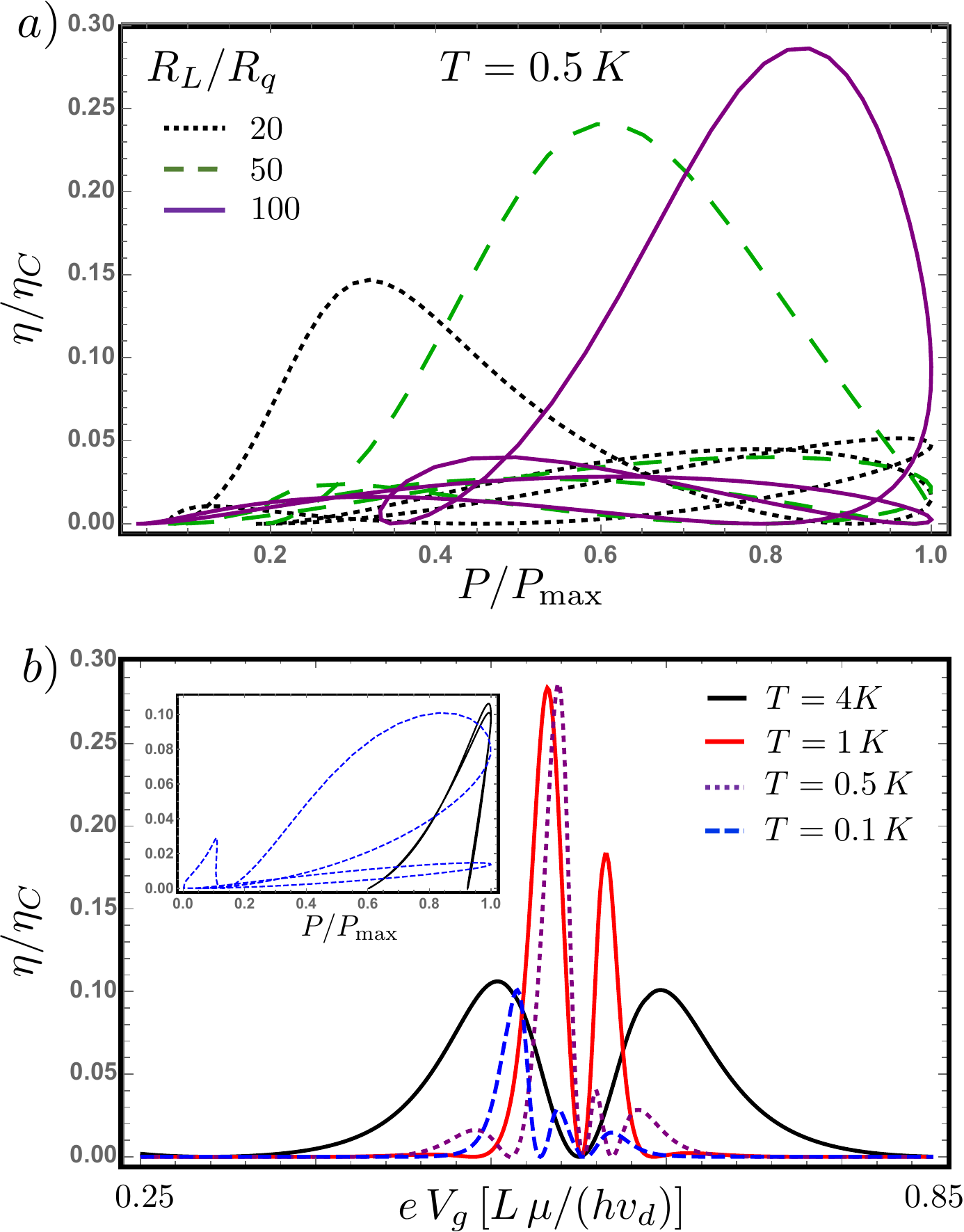}
\caption{a) Efficiency $\eta/\eta_C$ as a function of output power $P/P_\text{max}$ calculated at $T=0.5\,$K for a few values of load resistor $R_L$.
b) Efficiency vs $V_g$ calculated for different temperatures, and optimal load resistances (see text). 
The inset shows $\eta/\eta_C$ vs output power $P/P_\text{max}$ calculated at $T=0.1$K (blue dashed line) and 4K (black full line). Length and asymmetry of the ring are fixed to $2 \mu m$ and $\delta L/L = 0.3$.
}
\label{fig:efficiency}
\end{center}
\end{figure}

To harvest power from the engine, we close the circuit as shown in Fig. \ref{ring} a) with a load resistance $R_L$. 
This load will develop a thermo-voltage $V$ in the steady-state regime, which satisfies current conservation, $I_c = V/R_L$, where $I_c$ is the electrical thermocurrent given by Eq. \eqref{eq:Onsager} circulating in the circuit. The generated thermo-voltage takes the form $V = R_L G \, S \, \Delta T / (G R_L + 1)$, from which the outpout power generated by the AB ring can be directly calculated in terms of the transport and thermoelectric coefficients.  
In the linear response regime, the efficiency as a function of the power can be expressed in terms of the $ZT$ coefficient \cite{Benenti17}, and therefore depends on the load resistance $R_L$ through the output power:
\bb
\label{eq:eff}
\frac{\eta}{\eta_C}(R_L) = \frac{P_\text{out}(R_L)/P_\text{max}}{2 \left( 1 + 2/ZT - \sqrt{1-P_\text{out}(R_L)/P_\text{max}}\right)} \,.
\ee
Here $\eta_C = \Delta T / T$ is the Carnot efficiency in the linear response regime, and the maximum power $P_\text{max}= G S^2 \Delta T^2/4$ is achieved for a thermo-voltage being half the stopping voltage $V_s = S \Delta T$.

We have exploited Eq. \eqref{eq:eff} to find the optimal value of the load resistance for different temperatures, as shown in Fig. \ref{fig:efficiency} a) for $T = 0.5\,$K. 
We took the maximal values of $ZT$ and the corresponding values for $\Phi_{AB}$ and $V_g$ as shown in Fig. \ref{fig:ZT}. 
The optimal load resistances in units of the quantum resistance $R_q = h/2 e^2$ for $T = \{ 0.1, 0.5, 1, 4\}$\, K are respectively $R_L / R_q = \{100, 100, 74.9, 4.6\}$. 

\begin{figure*}[t!]
\label{fig:transport}
\begin{center}
\includegraphics[width=0.7 \paperwidth]{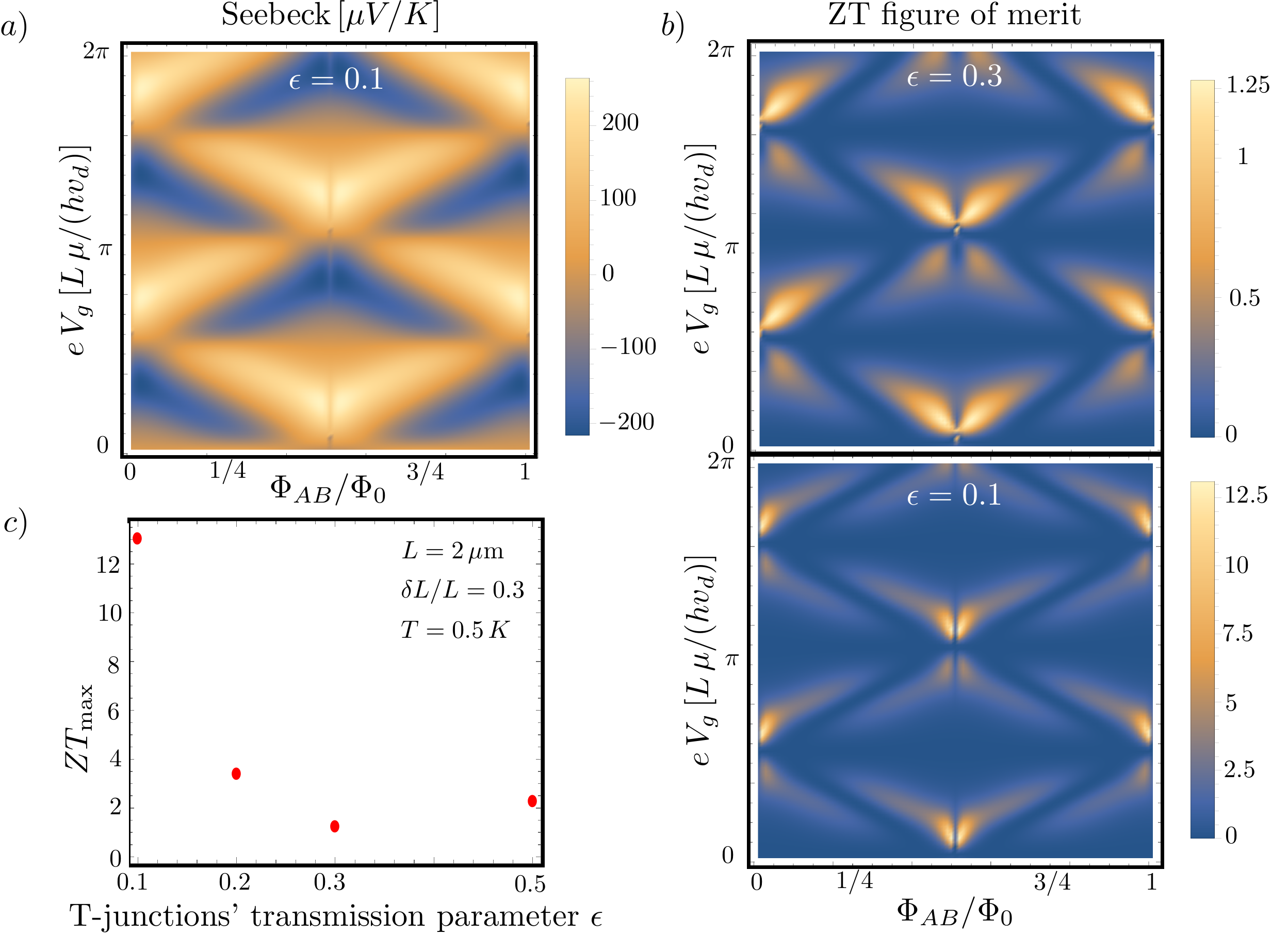}
\caption{Effect of finite transmission through the T-junctions on the thermoelectric response and figure of merit of the AB heat engine. We have fixed the length and asymmetry of the AB ring to $L=2 \mu m$ and $\delta L / L = 0.3$, and the temperature to $T=0.5 K$. a) Seebeck coefficient for $\epsilon = 0.1$. The maximum thermoelectric response is of the same order, $\sim 300 \mu K/V$ as for $\epsilon = 0.5$, see Fig. \ref{fig2}, but it exhibits a more pronounced sharp dependence on the gate voltage $V_g$ and AB flux $\Phi/\Phi_0$. This allows for achieving much higher values for the figure of merit, shown in panels b) and c). In b), we show density plots of the \emph{ZT} figure of merit for $\epsilon = 0.3$ and $\epsilon = 0.1$. For $\epsilon =0.3$, resonant tunneling does not yet take place, explaining lower \emph{ZT} values as compared to Fig. \ref{fig:ZT} where $ZT > 2$ for $L=2 \mu m, \delta L/L 0.3$ and $T=0.5$K. When transmission through the T-junctions is further decreased, $\epsilon=0.1$, resonant tunneling takes place and \emph{ZT} achieves exceptionally high values, exceeding 12. In panel c), we show the max values achieved for \emph{ZT} as a function of $\epsilon$ over the parameter space spanned by the gate voltage and the AB flux. It clearly shows the advantage provided by resonant tunneling once $\epsilon \leq 0.2$.}
\label{fig:5}
\end{center}
\end{figure*}

In the lower panel of Fig. \ref{fig:efficiency}, we show the efficiency $\eta/\eta_C$ from Eq. \eqref{eq:eff} as a function of the gate voltage $V_g$ for different temperatures and optimal load resistances. 
Remarkably, the efficiency reaches $28.3\%$ of Carnot efficiency for $T = 1\,$K and $28.6 \%$ at $T = 0.5\,$K.
It is also interesting to note that the efficiency reaches a similar maximum for the lowest and highest temperatures, $T=0.1$ and $T=4\,$K. 
However, the thermoelectric advantage of the AB heat engine at $T=0.1\,$K compared to $T=4\,$K is clearly demonstrated in the inset. 
Whereas the efficiency reaches its maximum at maximum power for $T=4$ K as expected when $ZT<1$, the maximal efficiency at $T=0.1\,$K does not correspond to the efficiency at maximum power.

At $T = 1\,$K and $T=0.5\,$K, the efficiency at maximum power reaches $\sim 26\%$ and $\sim 27\%$, respectively.
To the best of our knowledge, the highest efficiency at maximum power considering a mesoscopic engine has been recently obtained in Ref.~\onlinecite{Samuelsson17}, and reached $\sim 29\%$.
The corresponding device is an interferometer built within the quantum Hall regime, and the energy dependence of the transmission probability is induced by a time-dependent driven mesoscopic capacitor. From the experimental point of view, this device is more complex compared to a tunable AB ring, for a similar efficiency at maximum power. However, we would like to stress that these two proposals demonstrate the high potential of exploiting mesoscopic interferometers towards the realization of efficient quantum heat engines. 
The corresponding efficiencies at maximum power are close to the optimal one derived for a two-terminal engine, obtained with a box-car type transmission probability \cite{Whitney}.\\

\section{The AB heat engine under resonant tunneling conditions}
\label{Sec:T_junction}

In this last section, we wish to further emphasize the exceptional experimental tunabilities of the AB ring allowing for high thermoelectric performances when operating as a heat engine. This is done by exploiting the transmission probabilities of the T-junctions between the AB ring and the left and right leads. Following Ref. \onlinecite{Buttiker84} and already recalled in Sec. \ref{Sec:TAB}, this one can be characterized by a single parameter $\epsilon \in [0, 0.5]$ appearing in the total transmission probability $T_{AB}$, see eq. \eqref{eq:TAB}. When $\epsilon$ is small for both T-junctions, the transmission probability $T_{AB}$ exhibits resonances due to \emph{resonant tunneling} effect. This can be seen in Fig. \ref{ring} from the red curve, the transmission probability exhibits sharp peaks as a function of energy. Due to the other parameters (asymmetry, gate voltage), $T_{AB}$ remains highly asymmetric as a function of energy, allowing for exceptional thermoelectric properties. These ones are illustrated in Fig. \ref{fig:5}. For clarity, we have fixed the length, the asymmetry and the temperature to $L=2 \mu m, \delta L / L = 0.3$ and $T=0.5$K.

A lower transmission of the T-junctions characterized by $\epsilon <0.5$ leads for both the Seebeck coefficient and \emph{ZT} figure of merit to a much more pronounced dependence on the gate voltage $V_g$ and the AB flux as shown by the density plots in panels a) and b). However, an increase of the thermoelectric properties of the AB heat engine only arises when $\epsilon$ is sufficiently small, $\epsilon \leq 0.2$, such that resonant tunneling starts playing a role. This can be seen from panel b), where the \emph{ZT} values does not exceed 1.25 for $\epsilon = 0.3$, but reaches values as high as 12.5 when $\epsilon = 0.1$. 
Resonant tunneling is also put in evidence in panel c), where we show the maximum values obtained for \emph{ZT} over the full space spanned by $V_g$ and $\Phi_{AB}$ as a function of $\epsilon$. To the best of knowledge, high values for \emph{ZT} clearly exceeding the threshold 3 to have more than 30\% efficiency have only been predicted for very few systems so far, in engines exploiting superconducting gaps \cite{Ozaeta14}, quantum spin Hall states \cite{Roura18}, graphene doped by magnetic impurities \cite{Mani19} and mobility edges \cite{Chiaracane19}. Remarkably, the AB heat engine performance relies entirely (and solely) on single-particle quantum coherence, one of the simplest genuine quantum properties.

\section{Conclusions}

In summary, we have proposed and analyzed a phase-coherent mesoscopic heat engine based on a Aharonov-Bohm quantum interferometer.
The system can provide sizeable thermoelectric response, and large thermodynamic efficiency at maximum power ($\sim 30\%$) which is close to the optimal one achievable for a two-terminal heat engine.
Under conditions which are easily accessible from the experimental point of view, the heat engine is able to yield full phase and electrostatic control of thermal and electric conductance as well as of  its thermoelectric figures of merit.
 High-mobility GaAs/AlGaAs two-dimensional electron gas heterostructures \cite{Yamamoto12} are ideal candidates for the implementation of the AB quantum heat engine which is expected to lead to realistically robust performances over a wide range of system parameters configurations. Our results suggest the AB  interferometer as the archetypal quantum platform for the realization of unique phase-tunable heat engines and quantum thermal machines operating at cryogenic temperatures.\\

%\textit{Experimental considerations--}

\textit{Acknowledgements--} G.H. acknowledges support from the Swiss FNS through a starting grant PRIMA PR00P2$\_$179748. F.G. acknowledges the European Research
Council under the European Union’s Seventh Framework Programme (COMANCHE; European Research Council Grant No. 615187) and Horizon 2020 and innovation programme under grant agreement No. 800923-
SUPERTED. This research was supported in part  by the National Science Foundation under Grant No. NSF PHY17-48958 (KITP program QTHERMO18).

\appendix
\section{Derivation of the transmission probability through a AB ring}
\label{App:A}

\begin{figure}[t!]
\begin{center}
\includegraphics[width=0.5 \columnwidth]{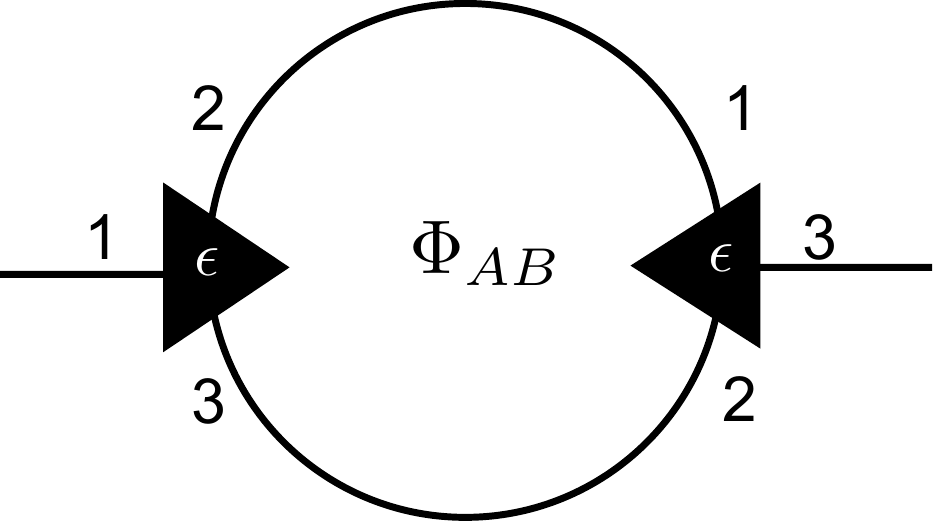}
\caption{Schematic of an AB ring in the presence of a magnetic field that leads to an enclosed magnetic flux $\Phi_{AB}$, with T-junctions (black traingles) characterized by a single parameter $\epsilon$.}
\label{fig:app}
\end{center}
\end{figure}

Here we provide details on the derivation of the transmission probability of the AB ring in full generality. We take into account finite transmission through the T-junctions characterized by the parameter $\epsilon$, the asymmetry of the AB ring and the presence of a magnetic field threading the sample that leads to an enclosed magnetic flux.  The schematic of the setup is shown in Fig. \ref{fig:app}. To satisfy symmetry between the two branches of the ring,  following Ref. \onlinecite{Buttiker84} and the labels of the different channels as shown on Fig. \ref{fig:app}, the real scattering matrices of the left and right T-junctions, $S_{t,L}$ and $S_{t,R}$ respectively, are given by :
\bb
S_{t,L} &=& \left( \begin{array}{ccc}
-\sqrt{1-2 \epsilon} & \sqrt{\epsilon} & \sqrt{\epsilon} \\
\sqrt{\epsilon} & a & b \\
\sqrt{\epsilon} & b & a \end{array} \right)  \equiv \left( \begin{array}{ccc}
r_{11}^L & t_{12}^L & t_{13}^L \\
t_{21}^L & r_{22}^L & t_{23}^L \\
t_{31}^L & t_{32}^L & r_{33}^L \end{array}\right)\,. \nonumber\\
%\ee
%\end{center}
%\begin{center}
%\bb
S_{t,R} &=& \left( \begin{array}{ccc}
a & b & \sqrt{\epsilon} \\
b & a & \sqrt{\epsilon} \\
\sqrt{\epsilon} &  \sqrt{\epsilon} & -\sqrt{1-2 \epsilon}  \end{array} \right) \equiv \left( \begin{array}{ccc}
r_{11}^R & t_{12}^R & t_{13}^R \\
t_{21}^R & r_{22}^R & t_{23}^R \\
t_{31}^R & t_{32}^R & r_{33}^R \end{array}\right)\,. \nonumber
\ee
with $a,b$ satisfying $(a+b)^2+ 2 \epsilon=1$ and $a^2+b^2 + \epsilon=1$ for current (probability) conservation:
\bb
a = \frac{1}{2} \left( \sqrt{1-2 \epsilon} -1\right) \quad \text{and} \quad b = \frac{1}{2} \left( \sqrt{1-2 \epsilon} +1\right)\,.
\ee
and $t_{ij}^{L,R}, r_{ij}^{L,R}$ the transmission and reflection amplitudes to go from input $j$ to output $i$ as defined in the left ($L$) and right ($R$)
 T-junctions. Then the total transmission amplitude takes into account contributions from paths starting from the upper arm or from the lower arm, that we denote $t_u$ and $t_d$ respectively following Ref. \onlinecite{Nazarov09}:
 \bb
 t = t_{21}^L t_u + t_{31}^L t_d\,.
\ee

Denoting by $\phi_i$ the phase acquired in presence of a perpendicular magnetic field along the upper ($i=1$) and lower ($i=2$) arms, and by $\chi_i = k_i L_i $ the dynamical phase where $k_i$ is the momentum vector and $L_i$ the length of arm $i$, the amplitudes $t_u, t_d$ can be shown to be the solutions of
\bb
t_u &=& e^{i (\phi_1+\chi_1)} t_{31}^R + e^{2 i \chi_1} (r_{11}^R r_{22}^L t_u + r_{11}^R t_{32}^L t_d) \nonumber \\
&& + e^{i (\chi_1 + \chi_2)} e^{i(\phi_1 - \phi_2)} (t_{21}^R t_{23}^L t_u + t_{21}^R r_{33}^L t_d)\,, \\
t_d &=& e^{i (\phi_2+\chi_2)} t_{32}^R + e^{2 i \chi_2} (r_{22}^R r_{33}^L t_d + r_{22}^R t_{23}^L t_u) \nonumber \\
&& + e^{i (\chi_1 + \chi_2)} e^{-i(\phi_1 - \phi_2)} (t_{12}^R t_{32}^L t_d + t_{12}^R r_{22}^L t_u)\,. 
\ee

Inserting the expressions of the transmission and reflection amplitudes as a function of the parameter $\epsilon$, we get
\bb
t &=& \sqrt{\epsilon} (t_u + t_d)
\ee
with $t_u$ and $t_d$ the solutions of
\bb
t_u &=& e^{i (\phi_1+\chi_1)} \sqrt{\epsilon} \nonumber \\
&&+ \frac{ e^{2 i \chi_1}+ e^{i (\chi_1 + \chi_2)} e^{i(\phi_1 - \phi_2)}}{4}  \Big[ (1-2 \epsilon) (t_u + t_d) + (t_u - t_d) \Big] \nonumber \\
&& +   \frac{ e^{2 i \chi_1} - e^{i (\chi_1 + \chi_2)} e^{i(\phi_1 - \phi_2)}}{2}\sqrt{1-2 \epsilon} \,t_u \,,\\
t_d &=& e^{i (\phi_2+\chi_2)} \sqrt{\epsilon} \nonumber \\
&& + \frac{e^{2 i \chi_2} + e^{i (\chi_1 + \chi_2)} e^{-i(\phi_1 - \phi_2)}}{4}  \Big[ (1-2 \epsilon) (t_u + t_d) - (t_u - t_d) \Big] \nonumber \\
&& +   \frac{e^{2 i \chi_2} - e^{i (\chi_1 + \chi_2)} e^{-i(\phi_1 - \phi_2)} }{2} \sqrt{1-2 \epsilon} \,t_d \,.
\ee
Straightforward algebra leads to a simple expression of the total transmission amplitude:
\begin{widetext}
\bb
t &=&\frac{ \epsilon \, e^{-i (\chi_1 + \chi_2)}  \Big( e^{i (\chi_1 + \phi_1)} + e^{i (\chi_2 + \phi_2)} - e^{i (2 \chi_1 + \chi_2 + \phi_2)} - e^{i(\chi_1 + 2 \chi_2 + \phi_1)} \Big)}{ \Big( \epsilon - 1 + \sqrt{1-2 \epsilon} \Big) \cos(\phi_1 - \phi_2) + \Big( \epsilon - 1 - \sqrt{1-2 \epsilon} \Big) \cos(\chi_1 - \chi_2) - 2 (\epsilon-1)  \cos(\chi_1 + \chi_2) - 2 i \epsilon  \sin(\chi_1 + \chi_2)   } \nonumber \\
&&
\ee
The total transmission probability for the AB ring as a function of $\epsilon$ takes the final form:
\bb
T_{AB} (\epsilon) &=& \frac{ 1 - \cos(\chi_1 + \chi_2) \cos(\chi_1 - \chi_2)  + \cos(\phi_1 - \phi_2) \big( \cos(\chi_1 - \chi_2) - \cos(\chi_1 + \chi_2) \big)}{ \sin^2(\chi_1 + \chi_2) + \Big( \frac{2 (1- \epsilon) \cos(\chi_1 + \chi_2) - (1-\epsilon - \sqrt{1-2 \epsilon}) \cos(\phi_1- \phi_2) - (1-\epsilon + \sqrt{1-2 \epsilon}) \cos(\chi_1- \chi_2)}{2 \epsilon} \Big)^2 } \,. \nonumber \\
&&
\ee
\end{widetext}
We remark that the final result only depends on the AB magnetic flux $\phi_1 - \phi_2 = 2 \pi \Phi_{AB}/\Phi_0$ and on the sum and difference of the dynamical phases $\chi = \chi_1 + \chi_2$ and $\delta \chi = \chi_1 - \chi_2$. This expression corresponds to Eq. \eqref{eq:TAB} in the main text. When $\epsilon = 1/2$ (fully transmitting T-junctions from the leads to the two branches), one recovers existing expressions in the literature.

\end{document}